\documentclass[
twocolumn,
superscriptaddress,
 amsmath,amssymb,
 aps,
]
{revtex4-2}

\usepackage[strict]{changepage}
\usepackage{graphicx}% Include figure files
\usepackage{dcolumn}% Align table columns on decimal point
\usepackage{bm}% bold math
\usepackage{units}

\usepackage{xr}
\usepackage{subfigure}
\usepackage{xcolor}
\usepackage{mathtools}
\usepackage{bbold}
\usepackage{placeins}
\usepackage{float}
\usepackage{empheq}

\usepackage[most]{tcolorbox}

\newcommand{\Pndagop}[2]{{{}\hat P_{}^{\vphantom{\dagger}}}\vphantom{P}_{#1}^{#2}}
\newcommand{\Pdagop}[2]{{{}\hat P_{}^{\dagger}}\vphantom{P}_{#1}^{#2}}

\makeatletter
\newcommand*{\addFileDependency}[1]{% argument=file name and extension
\typeout{(#1)}% latexmk will find this if $recorder=0
\@addtofilelist{#1}
\IfFileExists{#1}{}{\typeout{No file #1.}}
}\makeatother

\newcommand*{\myexternaldocument}[1]{%
\externaldocument{#1}%
\addFileDependency{#1.tex}%
\addFileDependency{#1.aux}%
}
\myexternaldocument{SI}

\begin{document}

\title{Ultrafast Optical Control of Rashba Interactions in a TMDC Heterostructure}

\author{Henry Mittenzwey}
\affiliation{Technische Universit\"at Berlin, Institut f\"ur Theoretische Physik, Nichtlineare Optik und Quantenelektronik, Hardenbergstra{\ss}e 36, 10623 Berlin, Germany}
\author{Abhijeet Kumar}
\author{Raghav Dhingra}
\affiliation{Freie Universität Berlin, Department of Physics, Arnimallee 14, 14195 Berlin, Germany}
\author{Kenji Watanabe}
\author{Takashi Taniguchi}
\affiliation{National Institute for Materials Science, 1-1 Namiki, Tsukuba 305-0044, Japan}
\author{Cornelius Gahl}
\author{Kirill I. Bolotin}
\affiliation{Freie Universität Berlin, Department of Physics, Arnimallee 14, 14195 Berlin, Germany}
\author{Malte Selig}
\author{Andreas Knorr}
\date{\today}
\affiliation{Technische Universit\"at Berlin, Institut f\"ur Theoretische Physik, Nichtlineare Optik und Quantenelektronik, Hardenbergstra{\ss}e 36, 10623 Berlin, Germany}

\date{\today}

\begin{abstract}
We investigate spin relaxation dynamics of interlayer excitons in a MoSe$_2$/MoS$_2$ heterostructure induced by the Rashba effect. In such a system, Rashba interactions arise from an out-of-plane electric field due to photo-generated interlayer excitons inducing a phonon-assisted intravalley spin relaxation. We develop a theoretical description based on a microscopic approach to quantify the magnitude of Rashba interactions and test these predictions via time-resolved Kerr rotation measurements. In agreement with the calculations, we find that the Rashba-induced intravalley spin mixing becomes the dominating spin relaxation channel above T = 50$\,$K. Our work identifies a previously unexplored spin-depolarization channel in heterostructures which can be used for ultrafast spin manipulation.
\end{abstract}
\maketitle

\textit{Introduction:}
The progress towards one of the main goals of semiconductor spintronics, ultrafast spin manipulation, requires a detailed understanding of spin relaxation processes \cite{vzutic2004spintronics}. Spin relaxation control is especially important in monolayers of transition metal dichalcogenides (TMDCs) where the spin is coupled to the valley degree of freedom that can be optically initialized and read out via circularly-polarized light \cite{xiao2012coupled, xu2014spin, mak2012control}. While the spin relaxation of an exciton primarily occurs via the Bir-Aronov-Pikus (BAP) mechanism \cite{bir1975spin,lechner2005spin,maialle1996spin,maialle1993exciton,article:THEORY_intervalley_exchange_Selig2020,article:THEORY_prr_Selig2019,qiu2015nonanalyticity}, strong spin-orbit interactions in two-dimensional semiconductors also drive spin flips via
Dresselhaus \cite{dresselhaus1955spin} or Rashba  mechanisms \cite{article:THEORY_DispersionBychkovRashba1984, d1971spin,bravo2023spin,vaughan2018modeling,silva1997exciton,eldridge2010rashba,eldridge2008all,ochoa2013spin,dery2015polarization,chen2020tunable,hu2018intrinsic,gupta2021dictates,lin2019rashba}, which can emerge in the Elliott-Yafet (EY) \cite{elliott1954theory,yafet1963g,song2013transport,ochoa2012elliot,huertas2009spin} or Dyakonov-Perel limit (DP) \cite{d1971spin,bravo2023spin,vaughan2018modeling,silva1997exciton}, depending on the electron-phonon scattering conditions \cite{boross2013unified}.
The Rashba mechanism, in particular, can be induced in mirror inversion-symmetric materials \cite{huertas2009spin} – such as TMDCs \cite{yang2020exciton,shimotani2013zeeman,wu2013electrical} – by an external out-of-plane electric field \cite{book:Spin_orbit_coupling_Winkler2003}:
\begin{align}
    H_{\text{BR}} = -\frac{\alpha_{\text{BR}}}{\hbar}\boldsymbol{\sigma}\cdot\left(\mathbf E(\mathbf r)\times\mathbf p\right).
    \label{eq:Hamiltonian_Rashba_General}
\end{align}
Here, $\mathbf E(\mathbf r)$ denotes an out-of-plane electric field, $\boldsymbol{\sigma}$ is the vector of the Pauli matrices, $\mathbf p$ is the electron momentum and $\alpha_{\text{BR}}$ is the coupling constant.
\newline
Due to the occurring electric field in Eq.~\eqref{eq:Hamiltonian_Rashba_General}, the Rashba mechanism is especially interesting in type-II heterostructures built from two dissimilar TMDCs. Here, an optical excitation generates interlayer excitons, quasiparticles composed of layer-separated electrons and holes. Critically, the interlayer excitons experience an out-of-plane electric field $\mathbf E(\mathbf r)$ acting in Eq.~\eqref{eq:Hamiltonian_Rashba_General} due to charge separation across the interface. Therefore, the Rashba spin interactions, Eq.~\eqref{eq:Hamiltonian_Rashba_General}, in these systems can be controlled by the photo-induced exciton density on a sub-picosecond timescale. Despite their critical role for the spin dynamics, the Rashba interactions in TMDC heterostructures have not been investigated until now.
\newline
In this combined theoretical/experimental study, we start by developing a theoretical model of the effects of photo-induced electric fields on spins in a MoSe$_2$/MoS$_2$ heterostructure. We evaluate the Rashba spin-orbit coupling of interlayer excitons, Eq.~\eqref{eq:Hamiltonian_Rashba_General}, and determine its influence on phonon-assisted spin/valley relaxation dynamics.\\
To compare our results to experiments, we carry out time-resolved Kerr rotation (TRKR) measurements. By controlling the electric field, Eq.~\eqref{eq:Hamiltonian_Rashba_General}, via femtosecond laser pulses and external gate electrodes, we demonstrate the Rashba-induced phonon-assisted spin relaxation mechanism in the MoSe$_2$/MoS$_2$ heterostructure. The optical control of the Rashba interactions demonstrated in this Letter establishes interlayer excitons as viable candidates for ultrafast optospintronics.
\newline
MoSe$_2$/MoS$_2$ heterostructures are ideal candidates to study Rashba interactions. A weak conduction band (CB) splitting in MoS$_2$ \cite{marinov2017resolving,article:THEORY_kp_theory_Kormanyos2015} causes a small energy separation between the two lowest-lying interlayer excitons. This, in turn, ensures efficient spin mixing between the two excitons via Rashba coupling (Eq.~\eqref{eq:Hamiltonian_Rashba_General}) \cite{article:THEORY_Bychkov_Rashba_Coupling_Kormanyos2014}. In contrast, in tungsten-based heterostructures, a large conduction band splitting renders the spin relaxation via Rashba interactions unobservable on sub-ns time scale.\\

\textit{Photo-induced interlayer electric field:}
\begin{figure}[h!]
\subfigure[]{
\includegraphics[width=0.3\linewidth]{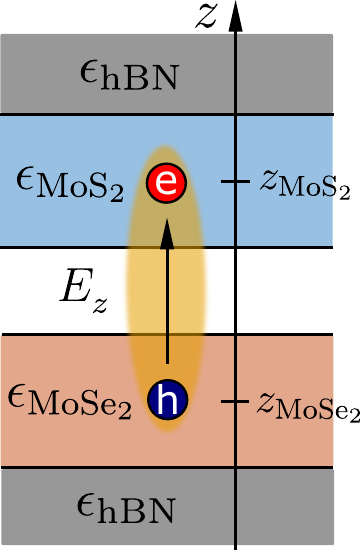}
\label{fig:MoSe2MoS2HSField}
}
\hspace{0.5cm}
\subfigure[]{\includegraphics[width=0.37\linewidth]{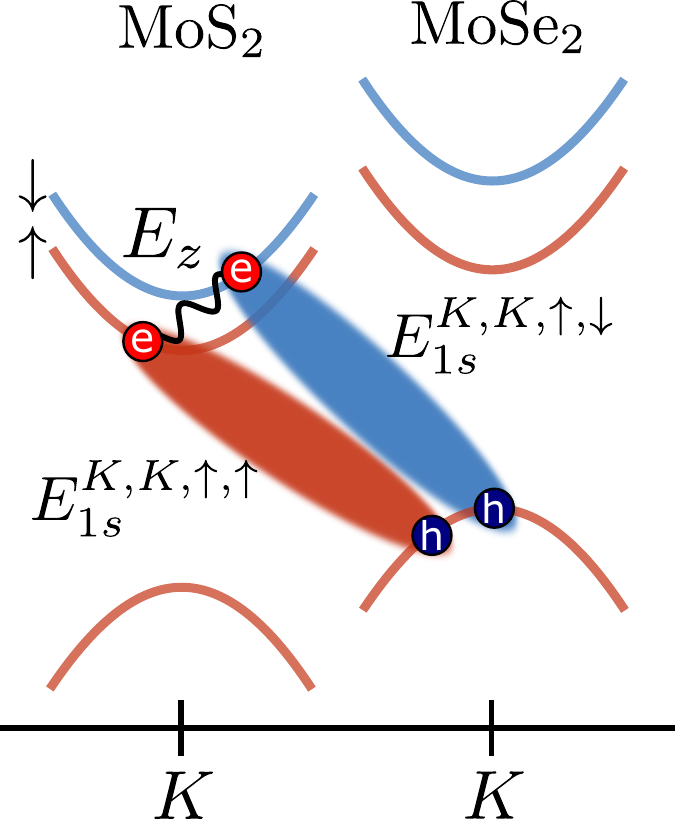}
\label{fig:rashba_excitons}
}
\caption{(a) Layer-separated excitonic interlayer occupations $N^{\text{IL}}$ (yellow-shaded ellipse) give rise to an out-of-plane electric field $E_z^{}$. $\epsilon_{\text{MoSe}_2/\text{MoS}_2/\text{hBN}}$ denotes the static dielectric constant of the respective layer (red-, blue-, and gray-shaded region) \cite{article:THEORY_laturia2018dielectric}, $z_{\text{MoSe}_2/\text{MoS}_2}$ is the $\delta$-confined position of the hole/electron.
(b) The two energetically lowest momentum-direct $1s$ interlayer excitons with energy $E_{1s}^{K,K,\uparrow,\uparrow}$ (spin-like, red) and $E_{1s}^{K,K,\uparrow,\downarrow}$ (spin-unlike, blue) are electron spin-coupled via $E_z$.
}
\end{figure}
To illustrate the concept of the spin relaxation mechanism, Fig.~\ref{fig:MoSe2MoS2HSField} displays the device geometry. Monolayers of MoS$_2$ (blue-shaded) and MoSe$_2$ (red-shaded) with a small vacuum gap in between are encapsulated in hBN (grey-shaded). The photo-induced  interlayer excitons are formed by charge transfer (yellow ellipse), consist of layer-separated holes in MoSe$_2$ ($z_{\text{MoSe}_2}$) and electrons in MoS$_2$ ($z_{\text{MoS}_2}$) and induce a real space charge density separation. This charge density causes a spatially homogeneous electric field, analog to the “capacitor formula” \cite{zimmermann2007exciton}, and depends on the dielectric environment $\epsilon_{\bot,\ell}$ and the total excitonic interlayer occupation $N^{\text{IL}}$, cf. Sec.~S1:
\begin{align}
    E_{z}^{\ell} = \frac{eN^{\text{IL}}_{}}{\epsilon_0\epsilon_{\bot,\ell}}.
    \label{eq:ElectricField}
\end{align}
Here, the layer index denotes $\ell=\{\text{MoSe}_2,\text{MoS}_2\}$, respectively, cf. Fig.~\ref{fig:MoSe2MoS2HSField}; The subscript "$\bot$" denotes the out-of-plane component of the static dielectric constant \cite{article:THEORY_laturia2018dielectric}; $e>0$ and $\epsilon_0$ are elementary charge and vacuum permittivity, respectively.
\newline
This electric field $E_{z}^{\ell}$, controlled by photo-induced interlayer exciton density $N^{\text{IL}}$, explicitly generates a Rashba spin-momentum coupling, cf. Eq.~\eqref{eq:Hamiltonian_Rashba_General}, and causes spin hybridization of interlayer excitons with different spins.
After the build-up of the interlayer excitons via charge transfer, $E_{z}^{\ell}$ is approximately static on timescales longer than sub-picosecond charge transfer \cite{policht2023time,schmitt2022formation,kumar2021spin,zheng2021thickness,article:EXP_charge_transfer_time_MoSe2WSe2_WangDalConte2021} and shorter than interlayer excitons radiative lifetime (tens of nanoseconds) \cite{article:EXP_THEORY_interlayer_excitons_direct_indirect_JahnkeWurstbauer2017,article:EXP_type2bandalign_Rivera2015,article:EXP_radiative_lifetimes_interlayer_vs_intralayer_Palummo2015}.

\textit{Spin-orbit coupling for interlayer excitons:}
In second quantization, the electronic Bychkov-Rashba Hamiltonian from Eq.~\eqref{eq:Hamiltonian_Rashba_General} for an out-of-plane electric field $\mathbf E = E_z^{\ell}\mathbf e_z$ from Eq.~\eqref{eq:ElectricField} translates to:
\begin{align}
    H_{\text{BR}} = \sum_{\substack{\mathbf k,\lambda,\xi,s,\ell}}
    S_{\mathbf k,\lambda}^{\xi,s,\ell}
    {a^{\dagger}_{}}\vphantom{a}_{\lambda,\mathbf k}^{\xi,s,\ell}{a^{\vphantom{\dagger}}_{}}\vphantom{a}_{\lambda,\mathbf k}^{\xi,\bar s,\ell},
    \label{eq:Hamiltonian_BR}
\end{align}
with the Rashba coupling matrix element, cf. Sec.~S2:
\begin{align}
    S_{\mathbf k,\lambda}^{\xi,s,\ell} = \alpha_{\text{BR}}^{\lambda,\xi,\ell}
    E_{z}^{\ell}\left(k_{y}-\mathrm i\left(\delta_{s,\uparrow}-\delta_{s,\downarrow}\right) k_{x}\right),
    \label{eq:spin_flip_matrix_element}
\end{align}
for momentum $\mathbf k$, band $\lambda$, valley $\xi$, spin $s$ and layer $\ell$. $a^{(\dagger)}_i$ are the annihilation (creation) operators for electrons, where $i=\{\mathbf k_i,\lambda_i,\xi_i,s_i,\ell_i\}$ is a compound index. The Rashba Hamiltonian, Eq.~\eqref{eq:Hamiltonian_BR}, couples different hole ($\lambda=v$) and electron ($\lambda=c$) spins $s,\bar s$ (if $s=\uparrow$ then $\bar s=\downarrow$ and vice versa) due to the out-of-plane electric field $E_z^{\ell}$ in Eq.~\eqref{eq:ElectricField}, cf. Fig.~\ref{fig:rashba_excitons}.

\begin{figure}[h!]
\begin{tabular}{cc}
\subfigure[]{\includegraphics[width=0.5\linewidth]{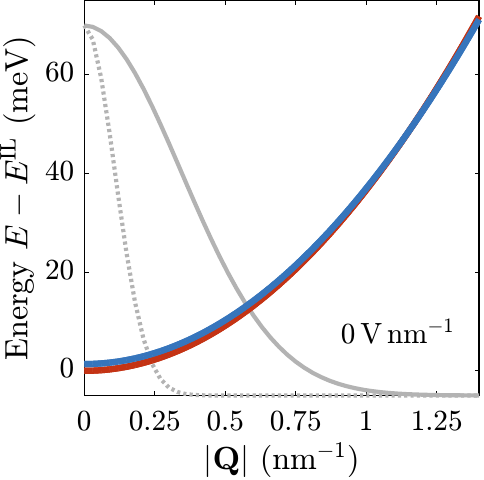}
\label{fig:RashbaDispersionE_z0}
}
\subfigure[]{\includegraphics[width=0.47\linewidth]{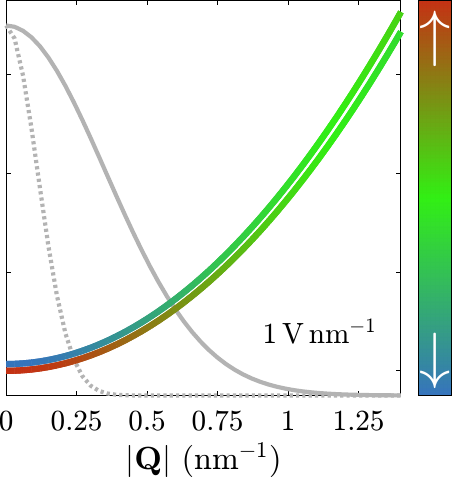}
\label{fig:RashbaDispersionE_z1}
%\vspace{-4cm}
}
\end{tabular}
\caption{
Excitonic dispersion with Rashba coupling from Eq.~\eqref{eq:Hamiltonian_BR} for the two energetically lowest interlayer exciton states, the spin-like exciton ($E_{1s,\mathbf Q}^{K,K,\uparrow,\uparrow}$), cf. red ellipse in Fig.~\ref{fig:rashba_excitons} and the spin-unlike exciton ($E_{1s,\mathbf Q}^{K,K,\uparrow,\downarrow}$), cf. blue ellipse in Fig.~\ref{fig:rashba_excitons}, at $E_{z} = 0\,\text{V}\,\text{nm}^{-1}$ (a) and $1\,\text{V}\,\text{nm}^{-1}$ (b). Red coloring denotes contribution of the spin up state ($\uparrow$), blue coloring the spin down state ($\downarrow$) of the electron and green coloring denotes the degree of spin mixing. The grey solid (dotted) lines denote the (normalized) Boltzmann distribution of the energetically lowest interlayer exciton at a temperature of 100$\,$K (10$\,$K).
}
\label{fig:RashbaDispersions}
%\end{center}
\end{figure}

\textit{Spin-diagonal exciton Hamiltonian:}
To establish the diagonalized exciton dispersion necessary to determine scattering events in the proper basis,
we transform Eq.~\eqref{eq:Hamiltonian_BR} into the excitonic picture \cite{article:THEORY_Katsch_pssb2018}, diagonalize the excitonic Rashba Hamiltonian with respect to the spin quantum numbers $s,\bar s$, cf. Sec.~S3. All band structure parameters are obtained from Ref.~\cite{article:THEORY_kp_theory_Kormanyos2015}, other parameters regarding the sample structure are provided in Tab.~S1.\\
It turns out, that the calculated spin hybridization is comparably weak for realistic photo-induced electric fields in the relevant momentum range of exciton-phonon scattering, cf. Fig.~\ref{fig:RashbaDispersions}. Therefore, from now on, we refer to all excitonic states with respect to the old spin basis $s=\{\uparrow,\downarrow\}$.
This ensures a more transparent discussion while no significant information is lost.\\
\\
In Fig.~\ref{fig:RashbaDispersions}, we show the spin-diagonalized dispersion $E_{1s,\mathbf Q}^{K,K,s,s^{\prime}}$
of the two energetically lowest momentum-direct interlayer excitons at the $K$ valleys as a function of momentum with hole/electron spin $s$/$s^{\prime}$, cf. Fig.~\ref{fig:rashba_excitons}. The plotted range is restricted to the center-of-mass momentum range $\mathbf Q$ relevant for the energetically lowest thermalized interlayer occupations at $T=100\,$K, cf. grey solid lines.
The dispersions at zero electric field are shown in Fig.~\ref{fig:RashbaDispersions}(a):
At zero center-of-mass momentum, the spin-like exciton is energetically lower than the spin-unlike interlayer exciton, since short-range intravalley exchange \cite{qiu2015nonanalyticity,robert2020measurement} is expected to be negligible in a heterobilayer. We note, that for a complete picture density-dependent energy shifts also have to be considered \cite{steinhoff2023exciton,erkensten2021exciton,schindler2008analysis,zimmermann2007exciton}.
Due to its larger effective mass,
the dispersion of the spin-like exciton eventually crosses the spin-unlike exciton at high center-of-mass momenta $|\mathbf Q|\approx 1.2\,\text{nm}^{-1}$.
In Fig.~\ref{fig:RashbaDispersions}(b), the dispersions under a very high non-zero electric field of $E_z = 1\,\text{V}\,\text{nm}^{-1}$, which serves to overemphasize the effect, are displayed. Here, the spins of the electrons as part of the exciton mix
and an anti-crossing behavior is observed.
The effective hybridization and band repulsion result from the interplay of two effects:\\
(i) Since the Rashba coupling is linear in the electron momentum, cf. Eq.~\eqref{eq:spin_flip_matrix_element}, which carries over to the excitonic center-of-mass momentum $\mathbf Q$ in Eq.~(S37), spin mixing vanishes at zero center-of-mass momentum
and is more effective at higher momenta, cf. Fig.~\ref{fig:RashbaDispersions}(b).
For that matter, the spin hybridization in the momentum range relevant for thermalized excitonic occupations at a temperature of 100$\,$K (solid lines in Fig.~\ref{fig:RashbaDispersions}) is much stronger compared to 10$\,$K (dotted lines in Fig.~\ref{fig:RashbaDispersions}).
(ii) At the same time, the hybridization strongly depends on the relative energetic position of the involved excitonic states, i.e. hybridization is strongest at energy-degenerate states.
The discussed and diagonalized excitonic dispersion now provides the energies for the exciton-phonon interaction as the dominant scattering process in TMDCs above cryogenic temperatures \cite{katzer2023exciton,article:THEORY_dark_exciton_formation_Selig2018,article:THEORY_exciton_cascade_Brem2018,article:THEORY_excitonic_linewidth_Selig2016,article:THEORY_excitonic_basis_equations_methods_Thraenhardt2000}.

\textit{Exciton-phonon coupling:}
By employing the exciton-phonon interaction in the spin-diagonal basis, cf. Eq.~(S44), the exciton kinetics for the interlayer excitonic occupations
%$N{\vphantom{N}}_{\mathbf Q}^{\xi,\xi^{\prime},\mathcal S}$ in the new spin basis
$N{\vphantom{N}}_{\mathbf Q}^{i}$:
\begin{align}
    N{\vphantom{N}}_{\mathbf Q}^{i} = \left \langle \Pdagop{\mathbf Q}{i} \Pndagop{\mathbf Q}{i} \right \rangle,
    %N{\vphantom{N}}_{\mathbf Q}^{\xi,\xi^{\prime},\mathcal S} = \left \langle \Pdagop{\mathbf Q}{\xi,\xi^{\prime},\mathcal{S}} \Pndagop{\mathbf Q}{\xi,\xi^{\prime},\mathcal{S}} \right \rangle,
    %= \sum_{s,s^{\prime}}\Big|A^{\mathcal S,\ell,\ell^{\prime},\xi,\xi^{\prime}}_{s,s^{\prime},\mu,\mathbf Q}\Big|^2N_{\mu,\mathbf Q}^{\ell,\ell^{\prime},\xi,\xi^{\prime},s,s^{\prime}}
    \label{eq:N_SH_Definition}
\end{align}
can be derived \cite{article:THEORY_excitonic_basis_equations_methods_Thraenhardt2000,article:THEORY_correlation_expansion_Fricke1996}. Eq.~\eqref{eq:N_SH_Definition} accounts only for incoherent excitons after coherence decay \cite{article:THEORY_dark_exciton_formation_Selig2018}. Here, $i$ is a compound index, which fully characterizes the excitonic state with respect to valley and spin.
Applying a Born-Markov approximation \cite{katzer2023exciton,book:graphene_carbon_nanotubes_malic2013}, the corresponding Boltzmann equation reads:
\begin{align}
\begin{split}
    %\partial_tN{\vphantom{N}}_{\mathbf Q}^{\xi,\xi^{\prime},\mathcal{S}} = &\, \sum_{\substack{\mathbf K,\mathcal S^{\prime},\xi^{\prime\prime},\xi^{\prime\prime\prime}}}\left(\Gamma^{in,\xi^{\prime\prime},\xi^{\prime\prime\prime},\xi,\xi^{\prime}}_{\mathbf K,\mathbf Q,\mathcal S^{\prime},\mathcal S}N{\vphantom{N}}_{\mathbf K}^{\xi^{\prime\prime},\xi^{\prime\prime\prime},\mathcal S^{\prime}}\right.\\
    %&\left.- \Gamma^{out,\xi^{\prime\prime},\xi^{\prime\prime\prime},\xi,\xi^{\prime}}_{\mathbf K,\mathbf Q,\mathcal S^{\prime},\mathcal S}N{\vphantom{N}}_{\mathbf Q}^{\xi,\xi^{\prime},\mathcal S}\right).
    \partial_tN{\vphantom{N}}_{\mathbf Q}^{i} = &\, \sum_{\substack{\mathbf K,j}}\left(\Gamma^{in,j,i}_{\mathbf K,\mathbf Q}N{\vphantom{N}}_{\mathbf K}^{j} - \Gamma^{out,j,i}_{\mathbf K,\mathbf Q}N{\vphantom{N}}_{\mathbf Q}^{i}\right).
    \end{split}
    \label{eq:Boltzmann_Equation_SpinHybridBasis}
\end{align}
In Eq.~\eqref{eq:Boltzmann_Equation_SpinHybridBasis}, $\Gamma^{in/out,j,i}_{\mathbf K,\mathbf Q}$ denote the phonon-assisted scattering rates between states $\mathbf Q,i$ and $\mathbf K,j$, which encode intra-
%($\xi^{\prime\prime}=\xi$ and $\xi^{\prime\prime\prime}=\xi^{\prime}$)
as well as intervalley
%($\xi^{\prime\prime}\neq\xi$ or $\xi^{\prime\prime\prime}\neq\xi^{\prime}$)
scattering of the corresponding holes and electrons forming the exciton. Due to Rashba coupling-induced spin mixing, phonon-assisted scattering in Eq.~\eqref{eq:Boltzmann_Equation_SpinHybridBasis} does not conserve the spin anymore, so that phonon-assisted spin relaxation becomes possible. Details of Eq.~\eqref{eq:Boltzmann_Equation_SpinHybridBasis} are provided in Sec.~S4.

\begin{figure*}[t]
% \begin{figure}
% \begin{minipage}{\dimexpr\paperwidth-4cm}
\subfigure[]{\includegraphics[width=0.23\linewidth]{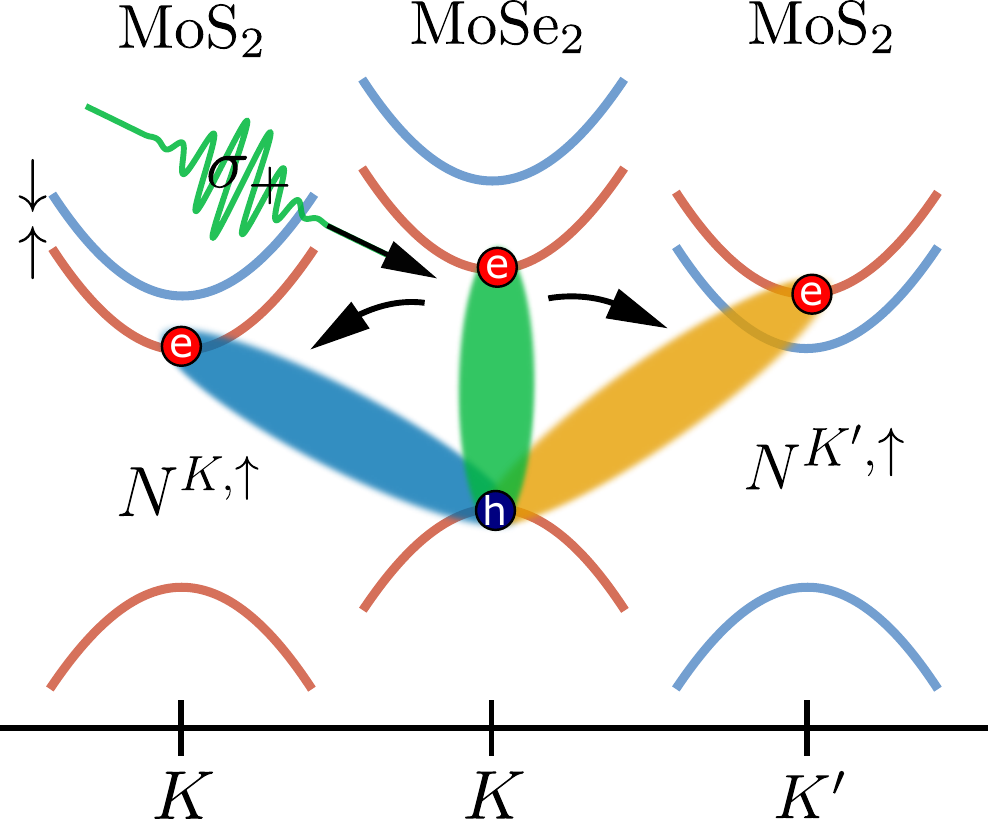}
\label{fig:pump_initial}
}
\subfigure[]{\includegraphics[width=0.23\linewidth]{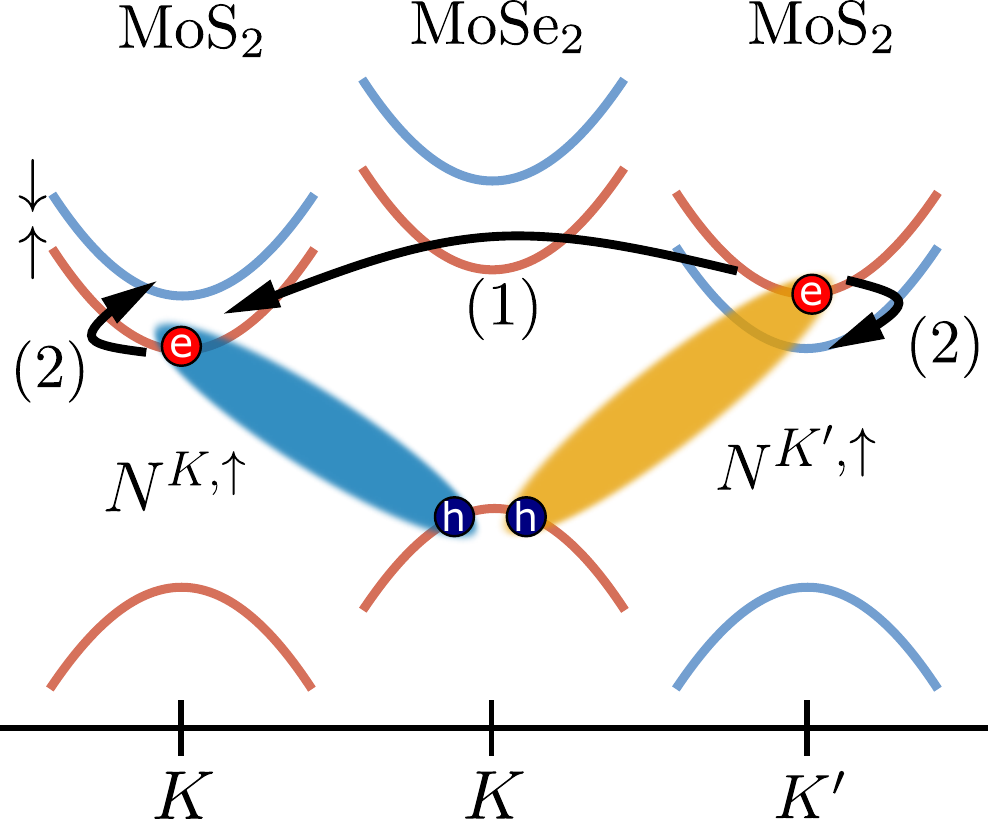}
\label{fig:scatteringdynamics}
}
\subfigure[]{\includegraphics[width=0.23\linewidth]{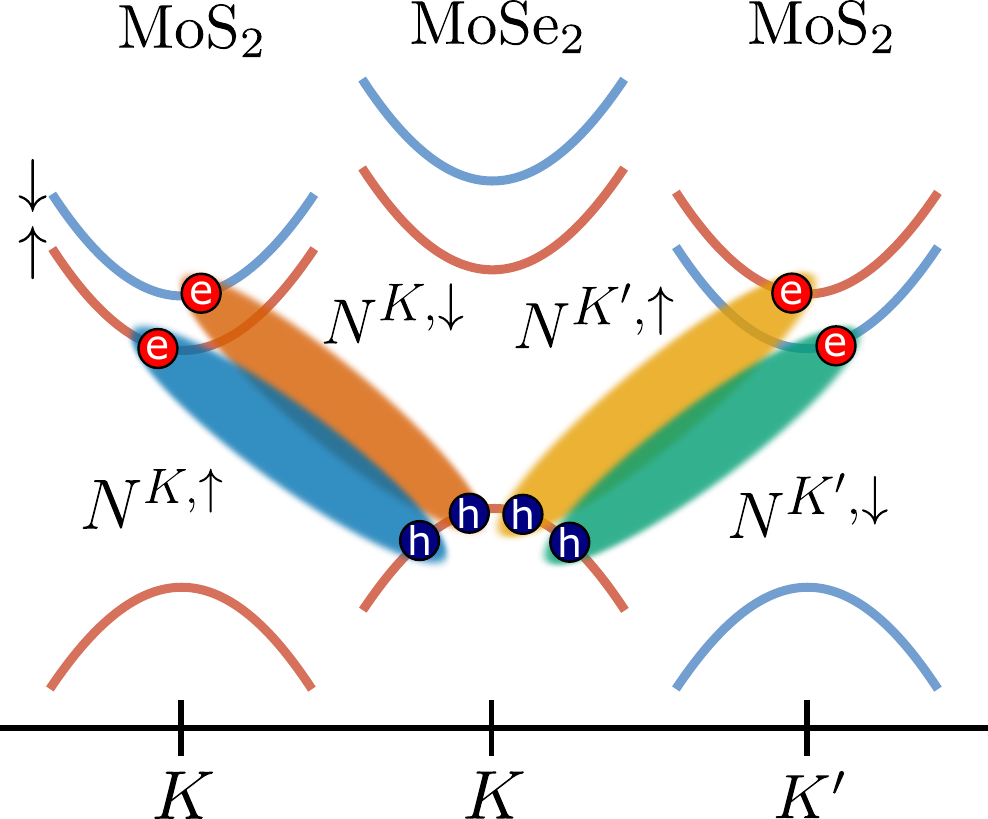}
\label{fig:equilibrium}
}
\subfigure[]{\includegraphics[width=0.23\linewidth]{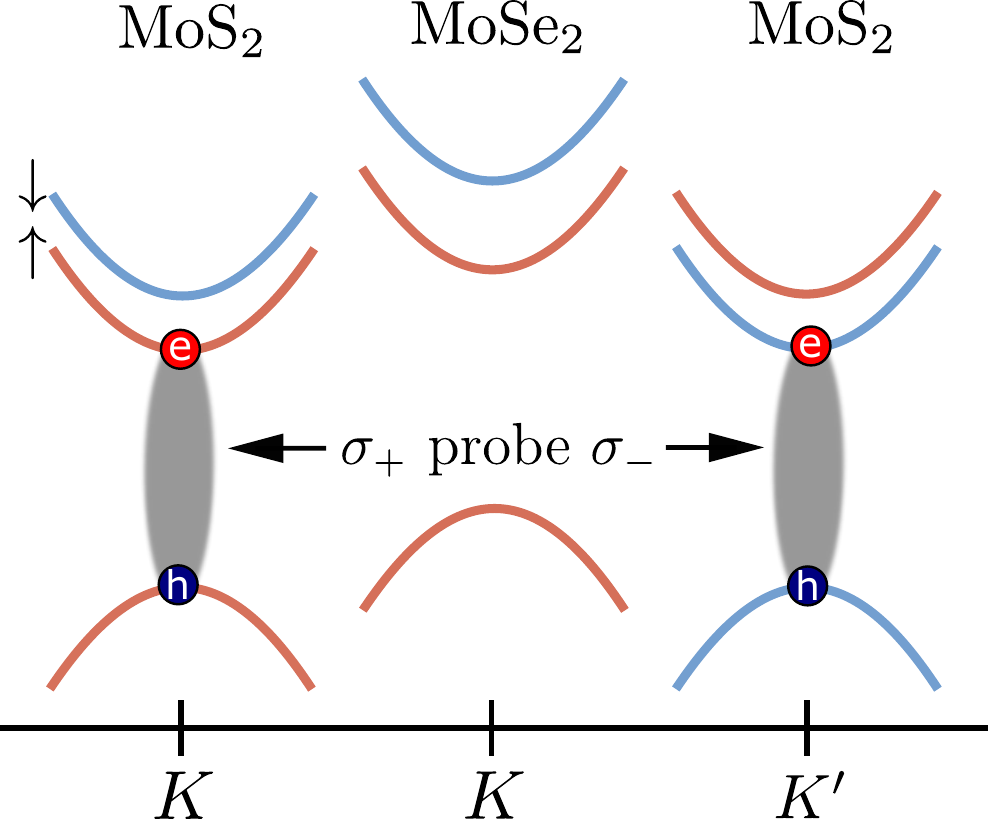}
\label{fig:probed_transitions}
}
\caption{The optically excited A$_{1s}$ transition by $\sigma_+$ light (a) in the MoSe$_2$ layer (green ellipse) leads to the formation of spin-like intra- ($N^{K,\uparrow}$, blue ellipse) and intervalley ($N^{K^{\prime},\uparrow}$, yellow ellipse) interlayer occupations following the ultrafast electron tunneling, which serve as initial conditions for the exciton kinetics in Eq.~\eqref{eq:Boltzmann_Equation_SpinHybridBasis}.
In (b), the most relevant relaxation processes encoded in Eq.~\eqref{eq:Boltzmann_Equation_SpinHybridBasis} are shown: (1) denotes spin-like phonon-assisted intervalley scattering and (2) denotes spin-unlike phonon-assisted intravalley scattering, which lead to an equilibrium of spin-like ($N^{K^{(\prime)},\uparrow}$) and spin-unlike ($N^{K^{(\prime)},\downarrow}$) excitonic occupations, as shown in (c). In (d), the TRKR dynamics are obtained via valley-selective excitation of the intralayer transitions in MoS$_2$ using the $\sigma_{+/-}$ components of a linearly-polarized probe.
}
\label{fig:ScatteringDynamics}
\end{figure*}

\textit{Exciton kinetics:}
We solve the Boltzmann equation in Eq.~\eqref{eq:Boltzmann_Equation_SpinHybridBasis} for four $1s$ interlayer excitonic species, i.e. the four lowest spin-like and spin-unlike intra- and intervalley interlayer excitons, cf. Fig.~\ref{fig:equilibrium},
in a moir\'e-free MoSe$_2$/MoS$_2$ heterostructure of $R_h^h$ stacking order \cite{article:THEORY_interlayer_excitons_Wozniak2020}.
Since the hole in MoSe$_2$ always remains spin-valley locked in the $K,\uparrow$-state, we label the excitonic occupations $N\vphantom{N}_{\mathbf Q}^i$ with respect to the valley and spin of the electron only.
The material parameters regarding phonon modes and deformation potentials are taken from Refs.~\cite{article:THEORY_phonon_parameters,li2013intrinsic}, where we take all relevant phonon modes (LO, TO, A$^{\prime}$, LA, TA) into account \cite{katzer2023impact}.
The initial conditions of our simulations are displayed in Fig.~\ref{fig:pump_initial}: We assume, that, on the timescale of the optical preparation (pumping the A$_{1s}$ resonance in the MoSe$_2$ layer, cf. green ellipse in Fig.~\ref{fig:pump_initial}) and the subsequent intra- and intervalley charge carrier tunneling ($\sim$ 100 femtoseconds) necessary for the interlayer exciton formation (indicated by the left and right arrows in Fig.~\ref{fig:pump_initial}, respectively) \cite{zheng2021thickness,kumar2021spin}, no spin flip occurs. Therefore, spin-like Boltzmann distributions for intravalley $K,K$ (blue ellipse in Fig.~\ref{fig:pump_initial}) and intervalley $K,K^{\prime}$ excitonic occupations (yellow ellipse in Fig.~\ref{fig:pump_initial}) are assumed as initial conditions of Eq.~\eqref{eq:Boltzmann_Equation_SpinHybridBasis}.
To examine the overall relaxation dynamics, we calculate from Eq.~\eqref{eq:Boltzmann_Equation_SpinHybridBasis} the total valley- and spin-resolved excitonic occupations $N{\vphantom{N}}_{}^{i} = \frac{1}{A}\sum_{\mathbf Q}N{\vphantom{N}}_{\mathbf Q}^{i}$,
where $A$ is the device area.
\begin{figure}[h!]
\includegraphics[width=1\linewidth]{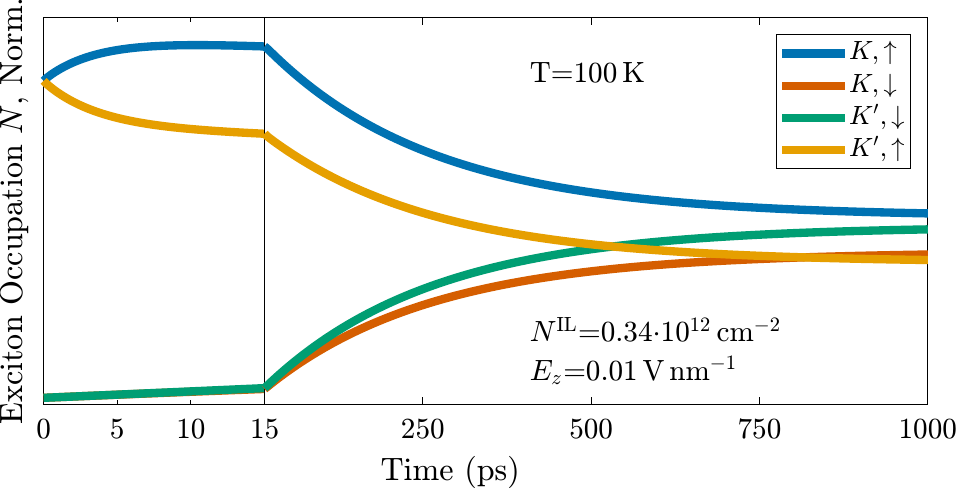}
\caption{Dynamics of the total excitonic occupations $N{\vphantom{N}}_{}^{K/K^{\prime},\uparrow/\downarrow}$ in Fig.~\ref{fig:ScatteringDynamics}(a)--(c) as solutions of Eq.~\eqref{eq:Boltzmann_Equation_SpinHybridBasis} at a temperature of 100$\,$K and a total interlayer exciton density of $3.4\cdot 10^{11}\,\text{cm}^{-2}$ with respect to the electron valley $K/K^{\prime}$ and spin $\uparrow/\downarrow$.
}
\label{fig:ExcitonicOccupation}
\end{figure}
In Fig.~\ref{fig:ExcitonicOccupation}, we display the dynamics of the four lowest interlayer occupations in Fig.~\ref{fig:equilibrium} as solutions of Eq.~\eqref{eq:Boltzmann_Equation_SpinHybridBasis}
at a lattice temperature of $100\,K$ and a total excitonic interlayer occupation over all valleys and spins of $N^{\text{IL}} = 3.4\cdot 10^{11}\,\text{cm}^{-2}$, which induces an electric field of $E_z = 0.01\,\text{V}\,\text{nm}^{-1}$, cf. Eq.~\eqref{eq:ElectricField}.
Initially ($t=0$) in Fig.~\ref{fig:ExcitonicOccupation}, the excitonic occupations $N{\vphantom{N}}_{}^{K,K,\uparrow,\uparrow}$ (intravalley, blue ellipse in Fig.~\ref{fig:pump_initial}) and $N{\vphantom{N}}_{}^{K,K^{\prime},\uparrow,\uparrow}$ (intervalley, yellow ellipse in Fig.~\ref{fig:pump_initial}) are populated.
Then, a two-step process is observed:
Within the first tens of picoseconds, intervalley phonon-assisted spin-like scattering leads to a decay of $N{\vphantom{N}}_{}^{K,K^{\prime},\uparrow,\uparrow}$
and a build-up of $N{\vphantom{N}}_{}^{K,K,\uparrow,\uparrow}$, cf. process (1) in Fig.~\ref{fig:scatteringdynamics}, until a temporarily equilibrium of spin-like occupations is reached.
On a timescale of several hundred picoseconds, spin-unlike phonon-assisted intravalley scattering caused by the photo-induced Rashba effect, cf. process (2) in Fig.~\ref{fig:scatteringdynamics}, becomes important and leads to a slow rise of intravalley (red solid line) and intervalley (green solid line) spin-unlike excitonic occupations. Simultaneously, a slow decay of the corresponding intravalley (blue line) and intervalley (yellow line) spin-like excitonic occupations occurs,
until an equilibrium at about $1\,\text{ns}$ is reached.

\textit{Time-resolved Kerr rotation (TRKR) signal:} The spin relaxation dynamics can be accessed by experimentally measuring the TRKR signal, $\theta_K(\tau)$, defined as the polarization rotation of the linearly-polarized probe upon reflection due to different absorption of the $\sigma_-$ and $\sigma_+$ components, cf. Fig.~\ref{fig:probed_transitions}.
This signal is approximately related to the imbalance of spin-like intravalley occupations $N^{K,\uparrow}$ (blue ellipse in Fig.~\ref{fig:ScatteringDynamics}) and spin-unlike intervalley occupations $N^{K^{\prime},\downarrow}$ (green ellipse in Fig.~\ref{fig:ScatteringDynamics}):
\begin{align}
\label{eq:KerrSignalOccupation}
    \theta_K(\tau) \sim N^{K,\uparrow}(\tau) - N^{K^{\prime},\downarrow}(\tau).
\end{align}
See Sec.~S5 for details.
\begin{figure}[h!]
\includegraphics[width=1\linewidth]{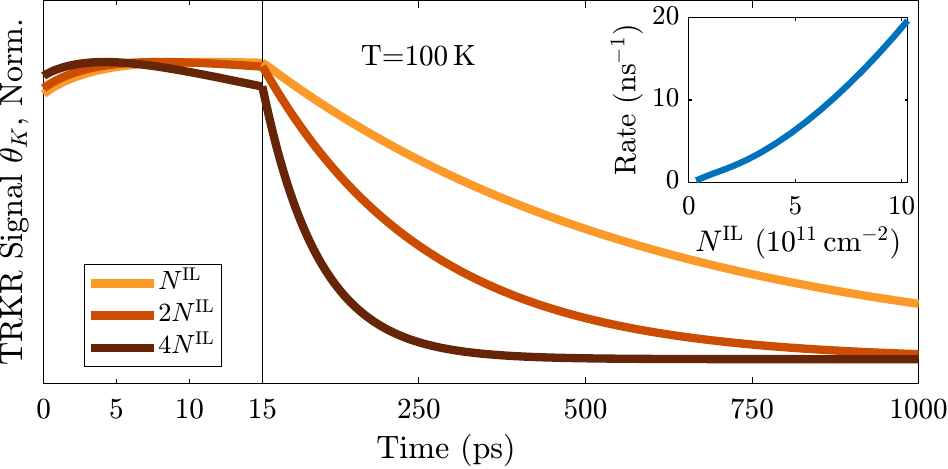}
\caption{Simulated TRKR timetraces for various interlayer exciton densities ($N^{\text{IL}} = 1.7\cdot 10^{11}\,\text{cm}^{-2}$), cf. Eq.~\eqref{eq:KerrSignalOccupation}. Inset: Interlayer exciton density-dependent decay rates extracted from the simulated TRKR timetraces.
}
\label{fig:KerrSignalDecayRate100KTheory}
\end{figure}
In Fig.~\ref{fig:KerrSignalDecayRate100KTheory}, we display the simulated TRKR timetraces from Eq.~\eqref{eq:KerrSignalOccupation} at various total interlayer densities $N^{\text{IL}}$ within the range of typical photo-induced interlayer exciton densities of $10^{11}\text{--}10^{12}\,\text{cm}^{-2}$.
Within the first tens of picoseconds, we observe a small initial build-up of the TRKR signal, which is related to phonon-assisted spin-like intervalley scattering from $N^{K^{\prime},\uparrow}$ (yellow ellipse in Fig.~\ref{fig:ScatteringDynamics}) to $N^{K,\uparrow}$ (blue ellipse in Fig.~\ref{fig:ScatteringDynamics}), cf. process (1) in Fig.~\ref{fig:scatteringdynamics}. On longer times, an overall decay is observed,
which reflects the equilibration of the spin-like intravalley occupation $N^{K,\uparrow}$ (blue ellipse in Fig.~\ref{fig:ScatteringDynamics}) and the spin-unlike intervalley occupation $N^{K^{\prime},\downarrow}$ (green ellipse in Fig.~\ref{fig:ScatteringDynamics}), due to phonon-assisted spin-unlike intravalley scattering, cf. process (2) in Fig.~\ref{fig:scatteringdynamics}.
Moreover, the overall TRKR decay time decreases with increasing total interlayer density $N^{\text{IL}}$, cf. orange to brown lines in Fig.~\ref{fig:KerrSignalDecayRate100KTheory}.
To quantify this effect, the extracted decay rates are shown in Fig.~\ref{fig:KerrSignalDecayRate100KTheory}(inset):
Within the considered range of interlayer exciton densities, we observe a nearly quadratic dependence of the TRKR decay rates on the total interlayer density $N^{\text{IL}}$.
Since charge-separated excitonic interlayer densities $N^{\text{IL}}$ induce out-of-plane electric fields, cf. Eq.~\eqref{eq:ElectricField}, which directly govern the Rashba coupling strength, cf. Eq.~\eqref{eq:Hamiltonian_BR}, and hence the degree of spin hybridization, cf. Fig.~\ref{fig:RashbaDispersions}, an increasing total interlayer density speeds up the spin relaxation and thus decreases the TRKR decay time.
By varying the pump fluence, which optically addresses the A$_{1s}$ transition in the MoSe$_2$ layer, the excitonic interlayer density is controlled, cf. Fig.~\ref{fig:pump_initial}. Thus, a strong fluence-dependence of a measured TRKR signal decay at 100$\,$K is expected.

\textit{Experiments:}
\begin{figure*}%[t!]
\includegraphics[width=0.85\linewidth]{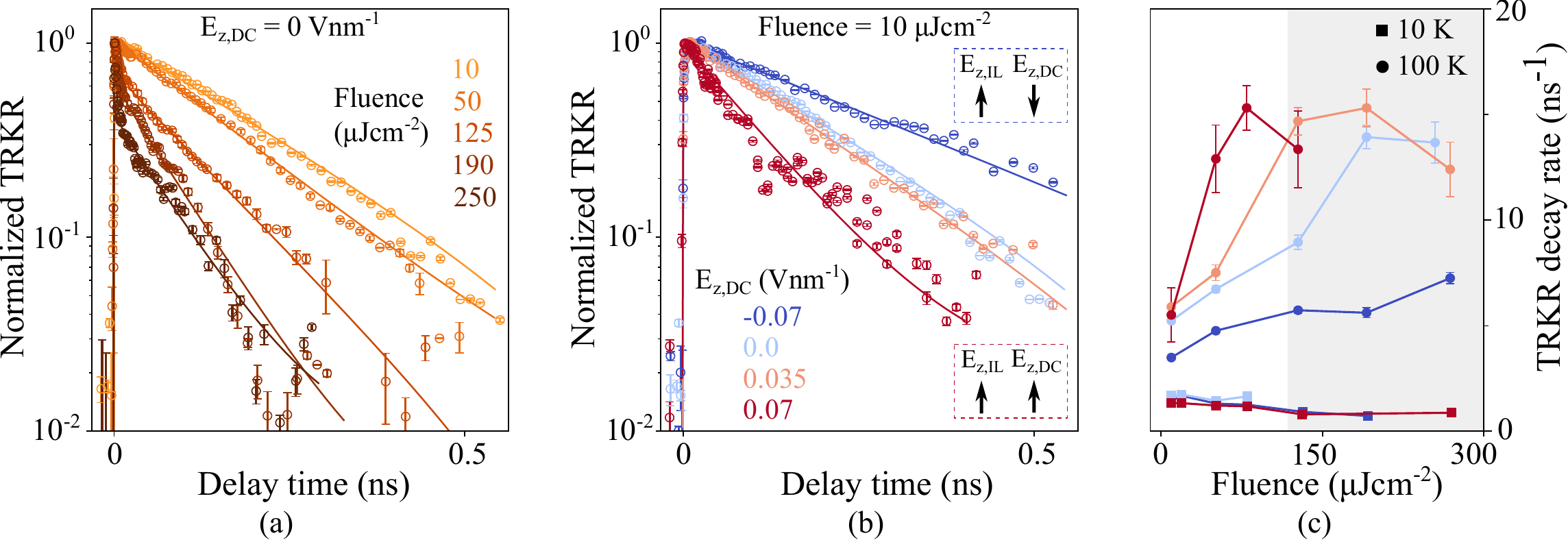}
\caption{(a) TRKR traces vs. pump fluence at zero external electric field and $T = 100\,$K. Solid lines are fit to the data. (b) TRKR traces vs. external electric field at a fluence of 10$\,$\textmu J$\,$${\text{cm}}^{-2}$. Arrows denote electric field direction. (c) TRKR decay rate vs. pump fluence for different values of $E_{z,\text{DC}}$ at 10$\,$K (square) and 100$\,$K (circle). The region corresponding to nonlinear fluence dependence of $N^{\text{IL}}$ is denoted by gray coloring.}
\label{fig:TRKR_vs_field_exp}
\end{figure*}
To test the theoretical scenario described above, we investigate double-gated MoSe$_2$/MoS$_2$ heterostructures via two-color TRKR microscopy with sub-200 fs time resolution (see Sec.~S9 and S10 for details). Here, a circularly-polarized pump pulse resonantly excites the MoSe$_2$ A$_{1s}$ resonance (1.65 eV), cf. Fig.~\ref{fig:pump_initial}, while a time-delayed linearly-polarized probe pulse (1.91 eV) probes the dynamics in the lower conduction sub-bands in MoS$_2$, cf Fig.~\ref{fig:probed_transitions}.
The electric field, $E_z$, in our device has two components: i) $E_{z,\text{DC}}$ due to the voltage difference between top and bottom gate electrodes, and ii) $E_{z,\text{IL}}$ due to the photo-induced charge density separation, cf. Fig.~\ref{fig:MoSe2MoS2HSField} and Eq.~\eqref{eq:ElectricField}. Note, that while $E_{z,\text{DC}}$ can be explicitly evaluated (see Sec.~S10 for details), the interlayer exciton density $N^{\text{IL}}$ is challenging to quantify due to optical saturation at elevated fluences and unknown charge transfer yield \cite{yagodkin2023probing}. However, there is always a monotonous relation between pump fluence and (interlayer) exciton density, so that we use pump fluence as a proxy to $E_{z,\text{IL}}$.\\
We begin by examining the TRKR traces vs. pump fluence at $E_{z,\text{DC}} = 0\,\text{V}\,\text{nm}^{-1}$ (Fig.~\ref{fig:TRKR_vs_field_exp}(a)) and vs. $E_{z,\text{DC}}$ at a low pump fluence of 10$\,$\textmu J$\,$${\text{cm}}^{-2}$ (Fig.~\ref{fig:TRKR_vs_field_exp}(b)) at $T = 100\,$K. Note the positive sign of $E_{z,\text{DC}}$ corresponds to the external field aligned along the interlayer exciton dipole. We find that the TRKR signal decays faster with increasing pump fluence. The TRKR signal also decays faster for $E_{z,\text{DC}} = 0.07\,\text{V}\,\text{nm}^{-1}$ compared to the signal for switched polarity, $E_{z,\text{DC}} = - 0.07\,\text{V}\,\text{nm}^{-1}$, cf. Fig.~\ref{fig:TRKR_vs_field_exp}(b). Overall, the decay accelerates when the $E_{z,\text{DC}}$ is aligned with the $E_{z,\text{IL}}$ dipole and when the fluence is large.
The observations in Fig.~\ref{fig:TRKR_vs_field_exp}(a),(b) are consistent with the spin relaxation being electric field-dependent, and establish pump fluence, i.e. total interlayer exciton density $N^{\text{IL}}$, as a control knob for Rashba-induced intravalley spin mixing.\\
To illustrate the efficiency of light-controlled Rashba interactions in our device, Fig.~\ref{fig:TRKR_vs_field_exp}(c) shows the experimentally obtained TRKR decay rate vs. pump fluence for different $E_{z,\text{DC}}$ at low ($T = 10\,$K, square) and high ($T = 100\,$K, circle) temperatures. We draw several conclusions from that data. First, a much weaker pump fluence and field dependence at 10$\,$K compared to 100$\,$K is consistent with a substantially lower rate of phonon-assisted spin-unlike scattering, cf. Fig.~S1. Second, consistent with Fig.~\ref{fig:TRKR_vs_field_exp}(a),(b), the rate of increase at 100$\,$K becomes superlinear when the $E_{z,\text{DC}}$ is aligned along the interlayer exciton dipole until a threshold fluence of 120$\,$\textmu J$\,$${\text{cm}}^{-2}$ is reached. A saturation beyond this value (gray area) is expected since $N^{\text{IL}}$ is influenced by nonlinear absorption, exciton bleaching, and entrance into the Mott screening regime \cite{article:EXP_low_density_limit_FoglerNovoselov2014}. Overall, the experimental results of Fig.~\ref{fig:TRKR_vs_field_exp} are in good qualitative agreement with theory predictions, cf. Fig.~\ref{fig:KerrSignalDecayRate100KTheory}.\\
We highlight the challenges inherent in comparing experimental observations with the theory, for details cf. Sec.~S12. First, as discussed earlier, the nonlinear dependence of the decay rate on pump fluence precludes quantitatively evaluating $E_{z,\text{IL}}$ and we have to rely on the directional comparison. Second, mechanisms other than Rashba relaxation (Auger \cite{yagodkin2023probing,steinhoff2021microscopic}, Coulomb scattering \cite{steinhoff2016nonequilibrium} and moiré-related processes \cite{brem2023bosonic,choi2021twist}) may contribute to the observed fluence dependence in Fig.~\ref{fig:TRKR_vs_field_exp}. Such mechanisms are hinted at by a weak but negative field dependence of the decay rate at $T = 10\,$K. Third, different electronegativities of the two layers and image charges may induce an additional electric field component \cite{barre2022optical,gupta2021dictates,li2017electric}, complicating the dynamics associated with the Rashba effect.

\textit{Conclusion:}
We proposed a new spin scattering mechanism of interlayer excitons in TMDC heterostructures with type-II band alignment. Charge-separated interlayer occupations induce an out-of-plane electric field, which mixes different spins due to the Rashba effect. This way, phonon-assisted scattering leads to spin relaxation at above cryogenic temperatures, and manifests in a time decay of the TRKR signal. By adjusting the pump fluence, we control the effective electric field leading to a new fluence dependent dynamics that is theoretically predicted and experimentally observed. Our findings have important consequences on the construction of spintronic devices utilizing type-II TMDC heterobilayers.\\

\textit{Acknowledgments:} H.M. and A. Knorr acknowledge financial support from the Deutsche Forschungsgemeinschaft (DFG) through Project KN 427/11-2 Project No. 420760124.
K.B. acknowledges the Deutsche
Forschungsgemeinschaft (DFG) for financial support
through the Collaborative Research Center TRR
227 Ultrafast Spin Dynamics (project B08). We thank D. Yagodkin, N. Stetzuhn and S. Kovalchuk for fruitful discussions.

\textit{Author contribution statements:} H.M. and A. Kumar contributed equally to this work. H.M. and M.S. did the theoretical work; A. Kumar and R.D. conducted the experiments with the help of C.G.

\FloatBarrier

\bibliography{bibliography.bib}
\end{document}